\documentclass[aps,prl,twocolumn,showpacs,preprintnumbers,amsmath,amssymb,superscriptaddress]{revtex4}
\usepackage{mathptm}  
\usepackage{dcolumn}                    
\usepackage{bm}                        
\usepackage{graphicx}
\usepackage{times}
\usepackage{epstopdf}
\begin{document}

\newcommand{\Imag}{{\Im\mathrm{m}}}   
\newcommand{\Real}{{\mathrm{Re}}}   
\newcommand{\im}{\mathrm{i}}        
\newcommand{\talpha}{\tilde{\alpha}}
\newcommand{\ve}[1]{{\boldsymbol{#1}}}

\newcommand{\x}{\lambda}  
\newcommand{\y}{\rho}     
\newcommand{\T}{\mathrm{T}}   
\newcommand{\Pv}{\mathcal{P}} 
\newcommand{\vk}{\ve{k}} 
\newcommand{\vp}{\ve{p}} 

\newcommand{\N}{\underline{\mathcal{N}}} 
\newcommand{\Nt}{\underline{\tilde{\mathcal{N}}}} 
\newcommand{\g}{\underline{\gamma}} 
\newcommand{\gt}{\underline{\tilde{\gamma}}} 

\newcommand{\vecr}{\ve{r}} 
\newcommand{\vq}{\ve{q}} 
\newcommand{\ca}[2][]{c_{#2}^{\vphantom{\dagger}#1}} 
\newcommand{\cc}[2][]{c_{#2}^{{\dagger}#1}}          
\newcommand{\da}[2][]{d_{#2}^{\vphantom{\dagger}#1}} 
\newcommand{\dc}[2][]{d_{#2}^{{\dagger}#1}}          
\newcommand{\ga}[2][]{\gamma_{#2}^{\vphantom{\dagger}#1}} 
\newcommand{\gc}[2][]{\gamma_{#2}^{{\dagger}#1}}          
\newcommand{\ea}[2][]{\eta_{#2}^{\vphantom{\dagger}#1}} 
\newcommand{\ec}[2][]{\eta_{#2}^{{\dagger}#1}}          
\newcommand{\su}{\uparrow}    
\newcommand{\sd}{\downarrow}  
\newcommand{\Tkp}[1]{T_{\vk\vp#1}}  
\newcommand{\muone}{\mu^{(1)}}      
\newcommand{\mutwo}{\mu^{(2)}}      
\newcommand{\epsk}{\varepsilon_\vk}
\newcommand{\epsp}{\varepsilon_\vp}
\newcommand{\e}[1]{\mathrm{e}^{#1}}
\newcommand{\dif}{\mathrm{d}} 
\newcommand{\diff}[2]{\frac{\dif #1}{\dif #2}}
\newcommand{\pdiff}[2]{\frac{\partial #1}{\partial #2}}
\newcommand{\mean}[1]{\langle#1\rangle}
\newcommand{\abs}[1]{|#1|}
\newcommand{\abss}[1]{|#1|^2}
\newcommand{\Sk}[1][\vk]{\ve{S}_{#1}}
\newcommand{\pauli}[1][\alpha\beta]{\boldsymbol{\sigma}_{#1}^{\vphantom{\dagger}}}

\newcommand{\eq}{Eq.}
\newcommand{\eqs}{Eqs.}
\newcommand{\cf}{\textit{cf. }}
\newcommand{\ie}{\textit{i.e. }}
\newcommand{\eg}{\textit{e.g. }}
\newcommand{\etal}{\emph{et al.}}
\def\i{\mathrm{i}}

\title{Majorana fermions manifested as interface-states in semiconductor hybrid structures}

\author{Jacob Linder}
\affiliation{Department of Physics, Norwegian University of
Science and Technology, N-7491 Trondheim, Norway}

\author{Asle Sudb{\o}}
\affiliation{Department of Physics, Norwegian University of
Science and Technology, N-7491 Trondheim, Norway}

\date{\today}

\begin{abstract}

Motivated by recent proposals for the generation of Majorana fermions in semiconducting hybrid structures, we  examine possible 
experimental fingerprints of such excitations. Whereas previous works mainly have focused on zero-energy states in vortex 
cores in this context, we  demonstrate analytically an alternative route to detection of Majorana excitations in semiconducting 
hybrid structures: interface-bound states that may be probed directly via conductance spectroscopy or STM-measurements. We 
estimate the necessary experimental parameters required for observation of our predictions.

\end{abstract}
\maketitle

The prediction \cite{bernevig_06, kane_prl_05} and experimental observation \cite{konig_jpsj_08, hsieh} of topological insulators 
has triggered an avalanche of research activity. Besides a number of fundamentally interesting aspects of the quantum spin Hall 
effect \cite{hirsch_prl_99} appearing in such systems, this class of materials also harbors a very real potential in terms of 
practical use in quantum computation. The reason for this is that they have been shown to host so-called Majorana fermions 
\cite{Majoranap} under a variety of circumstances \cite{akhmerov_prl_09, fu_prl_09, tanaka_prl_09, law_prl_09, linder_prl_10}. 
Such excitations satisfy non-Abelian statistics which form a centerpiece in recent proposals for topological quantum 
computations \cite{fu_prl_08}.

\begin{figure}[t!]
\centering
\resizebox{0.49\textwidth}{!}{
\includegraphics{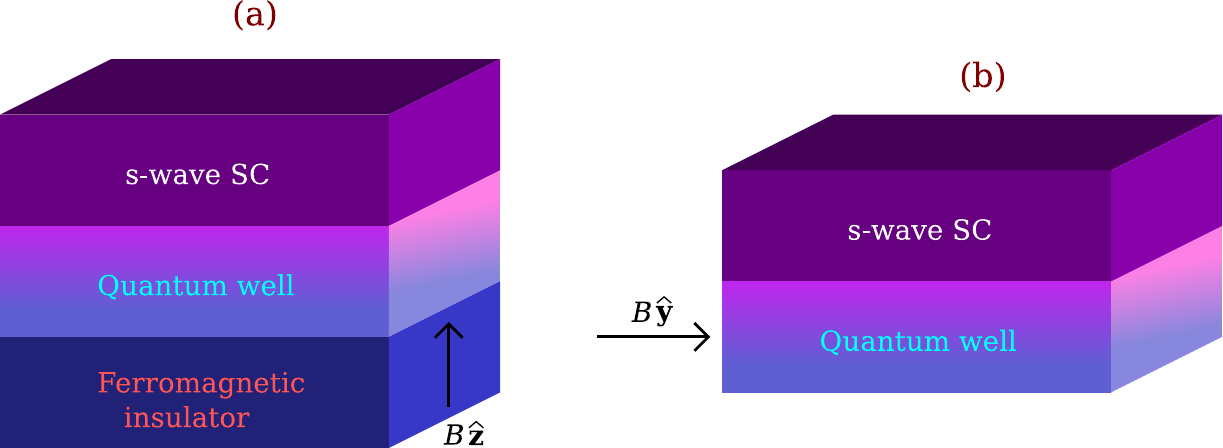}}
\caption{(Color online) The experimental setup proposed in Refs. \cite{sau_prl_10, alicea_prb_10} for generation of Majorana 
fermions in a semiconducting hybrid structure. In (a), a superconducting order parameter and Zeeman-interaction is induced by 
means of the proximity-effect in a quantum well with Rashba spin-orbit coupling, whereas in (b) the quantum well features a 
combination of Rashba and Dresselhaus spin-orbit coupling with an exchange interaction induced by an external field rather 
than a ferromagnetic insulator.
}
\label{fig:model} 
\end{figure}

From a technological point of view, the field of topological insulators is still in its infancy. Two recent works  
\cite{sau_prl_10, alicea_prb_10} that addressed the generation of Majorana fermions in \textit{semiconducting} devices 
have therefore attracted much attention, since semiconductor technology is very well-developed and thus offers greater 
experimental control over the system. The experimental setups suggested by Sau \etal\;\cite{sau_prl_10} and Alicea 
\cite{alicea_prb_10} are shown in Fig. \ref{fig:model}(a) and (b), respectively. Common for both proposals is that 
a quantum well with Rashba and/or Dresselhaus spin-orbit coupling is contacted to a superconducting reservoir and 
then driven into a topological phase by means of a magnetic field. When the latter exceeds a critical threshold, 
it effectively renders the band-structure in the quantum well formally equivalent to a spinless $k_x+\i k_y$ 
superconductor. This is a system which is known to host zero-energy Majorana fermions in vortex cores.

Up to now, it is precisely the prospect of Majorana fermions residing in vortex cores that has constituted the bulk 
of proposals for a realization of this exotic class of excitations in a condensed matter system. However, as we will 
show in this Letter, the Majorana fermions may also leave a distinct signature in semiconducting hybrid 
structures as the ones shown in Fig. \ref{fig:model}. Namely, interface-bound states with a unique dispersion which 
may be probed directly via conductance spectroscopy or STM-measurements. To demonstrate this, we will  first 
proceed to establish a direct correspondance between the systems considered in Fig. \ref{fig:model} and a spinless 
$k_x+\i k_y$ superconductor, and then calculate the energy dispersion for the interface-bound states analytically. 
The fingerprint of these states in STM-measurements would constitute a clearcut experimental observation of Majorana 
excitations in a condensed matter-system.

Both the presence of spin-orbit coupling and a Zeeman-interaction are key ingredients in establishing a topological 
superconducting phase in the systems suggested by Sau \etal~\cite{sau_prl_10} and Alicea \cite{alicea_prb_10}. The 
spin-orbit coupling ensures that a singlet-triplet mixing occurs for the induced superconducting order parameter, 
and thus generates a spinless $p$-wave order parameter. Upon introducing a Zeeman-field, one of the pseudospin-bands 
is raised above the Fermi level and one is left with a single-band spinless $p$-wave superconductor. Whereas such a 
Zeeman-field would have to be enormous in a conventional metal, the high $g$-factor and tunable Fermi level in 
semiconducting devices makes this possible even at fields below 1 T. An additional advantage of this is that the 
applied field then also remains well below the critical field $H_c$ for the proximity superconductor, which in many 
materials far exceeds 1 T \cite{sudbo_book}.

The purpose of this Letter is to demonstrate a clear experimental signature of the Majorana excitations proposed to exist in the 
setups of Refs. \cite{sau_prl_10, alicea_prb_10}, which also provides an alternative route to observation of Majorana fermions 
compared to the standard proposal of zero-energy vortex states. Our result applies both to Fig. \ref{fig:model}(a) and (b), but 
for the sake of not overburdening this work with analytical calculations we here focus on the setup in (a) which yields the most 
transparent results. The system in Fig. \ref{fig:model}(a) consists of \textit{i)} an $s$-wave superconductor, preferably with 
a high $T_c$ such as Nb, \textit{ii)} a quantum well semiconductor with Rashba spin-orbit coupling, such as InAs, and 
\textit{iii)} a ferromagnetic insulator such as EuO. The Hamiltonian for the conduction band of the quantum well then reads:
\begin{align}
\mathcal{H}_{QW} &= [-\nabla^2/(2m')-\mu]\hat{1} -\i\alpha(\partial_y\hat{\sigma}_x - \partial_x\hat{\sigma}_y),
\end{align}
where $\hat{\ldots}$ denotes a $2\times2$ matrix in spin space. Here, $m'$ is the effective mass of the electron (typically 
$m'\simeq m_e/20$), whereas $\alpha$ denotes the spin-orbit coupling constant. By means of the proximity effect to a 
ferromagnetic insulator, a Zeeman-field couples to the spins via:
\begin{align}
\mathcal{H}_{FI} &= -V_z\hat{\sigma}_z,
\end{align}
where $V_z$ is the magnitude of the exchange splitting. This interaction is strongly reduced compared to its value in the bulk 
ferromagnetic insulator, and it is thus reasonable to expect a magnitude of order $\mathcal{O}$(meV). The band-structure in 
the quantum well may now be obtained by diagonalizing the total Hamiltonian $\mathcal{H} = \mathcal{H}_{QW} + \mathcal{H}_{FI}$, 
which yields two pseudospin bands:
\begin{align}\label{eq:band}
\mathcal{E}^\beta_\vk = k^2/(2m') - \mu +\beta\sqrt{\alpha^2k^2 + V_z^2},\; \beta=\pm1.
\end{align}
Before introducing the superconducting proximity effect, it is instructive to pause briefly to consider the band-structure 
Eq. (\ref{eq:band}) in more detail. It follows that when the exchange interaction exceeds the chemical potential, $V_z>\mu$, 
the upper band is raised above the Fermi level for all momenta, i.e. $\mathcal{E}_\vk^+ > 0$. On the other hand, the lower 
band crosses the Fermi level at the momentum:
\begin{align}\label{eq:wavevector}
k_F = [2m'(m'\alpha^2 + \mu + \sqrt{m'\alpha^2(m'\alpha^2+2\mu) + V_z^2})]^{1/2}.
\end{align}
Enter now the superconducting pair field generated by the proximity $s$-wave superconductor. It adds a term to the Hamiltonian 
expressed by the original spinors $\psi = [\psi_\uparrow,\psi_\downarrow]$:
\begin{align}
\mathcal{H}_\text{SC} &= \int \text{d}^2\boldsymbol{r} [\Delta \psi_\uparrow^\dag(\vecr)\psi_\downarrow^\dag(\vecr) + \text{h.c.}]
\end{align}
Transforming the above equation into the new pseudospin basis of the long-lived excitations at Fermi level then produces the 
following gap for the lower band \cite{alicea_prb_10}
\begin{align}
\Delta_\vk = -\alpha\Delta(k_y-\i k_x)/(2\sqrt{V_z^2+\alpha^2k^2}).
\end{align}
The  Hamiltonian  can now be written
\begin{align}\label{eq:H}
\mathcal{H} &= \int \text{d}^2\vk \phi_\vk^\dag \mathcal{M}_\vk \phi_\vk,
\end{align}
where $\phi_\vk = [\varphi_\vk, \varphi_{-\vk}]$ is the pseudospin basis while 
\begin{align}
\mathcal{M}_\vk = \mathcal{E}_\vk \hat{\sigma}_z +\Delta_\vk\hat{\sigma}_x.
\end{align}
Here, we have defined $\mathcal{E}_\vk \equiv \mathcal{E}_\vk^-$ and the $\hat{\sigma}_j$ matrices now operate in pseudospin space. 

At this point, we can formally identify the obtained Hamiltonian as fully equivalent to a spinless $k_x+\i k_y$ superconductor 
(after a gauge transformation of $\e{\i\pi/2}$). We now proceed to demonstrate that the Majorana states in this system leave a 
unique fingerprint not only as zero-energy states in a vortex core, but also as \textit{interface-bound states}. Presumably, 
this simplifies greatly their experimental detection since one avoids the need to generate vortices in the quantum well. Instead, 
it suffices to probe the surface DOS at the edge of the quantum well either via conductance spectroscopy or STM-measurements. To be 
definite, let us consider the edge defined by $x=0$ (although our results are qualitatively identical for the edge $y=0$). 
Starting from the Hamiltonian Eq. (\ref{eq:H}), we construct the wavefunction in the quantum well which at $x=0$ takes the form
\begin{align}
\Psi(x=0) &= c_1\begin{pmatrix}
u_\vk\\
v_\vk \e{-\i\gamma_\vk^+} 
\end{pmatrix} + 
c_2 \begin{pmatrix}
v_\vk\e{\i\gamma_\vk^-}\\
u_\vk\\
\end{pmatrix},
\end{align}
where we have defined 
\begin{align}
\e{\i\gamma_\vk^\pm} = -(k_y\mp\i k_x)/k_F
\end{align}
and $u_\vk/v_\vk = \e{\i\text{acos}(\varepsilon/|\Delta_\vk|)}$. The constants $\{c_1,c_2\}$ are unknown and must be determined by 
proper boundary conditions. At the vacuum edge $x=0$, the wavefunction must vanish and we thus demand $\Psi(x=0)=0$, which allows for 
a determination of $\{c_1,c_2\}$. Doing so, we find that a non-trivial solution is obtained if the criterion 
\begin{align}
\left| \begin{array}{cc}
\e{\i\beta} & \e{\i\gamma_\vk^-} \\
\e{-\i\gamma_\vk^+} & \e{\i\beta} \\
\end{array}
\right| = 0
\end{align}
may be satisfied. This is indeed the case when:
\begin{align}\label{eq:bound}
|\varepsilon/\Delta| = \frac{\alpha k_F |\sin\theta|}{2\sqrt{V_z^2+\alpha^2k_F^2}},
\end{align}
where $k_F$ was defined previously. This equation describes precisely the announced interface-bound states and is one of the main results 
in this Letter. In general, subgap resonant-states are manifested as an enhanced DOS/peak-structure in such measurements whereas the rest 
of the subgap DOS would be suppressed due to the fully gapped Fermi surface. An important point to note is that since the present interface-state 
in Eq. (\ref{eq:bound}) is strongly dependent on the angle of incidence relative the edge, one would expect that the DOS to be enhanced in 
large parts of the subgap regime rather than featuring sharp spikes at isolated energies. Qualitatively, this would be experimentally seen 
as a broad hump-like structure in the conductance or surface DOS, similarly to the proposed chiral $p$-wave state in Sr$_2$RuO$_4$ \cite{mackenzie_rmp_03}.

\begin{figure}[t!]
\centering
\resizebox{0.49\textwidth}{!}{
\includegraphics{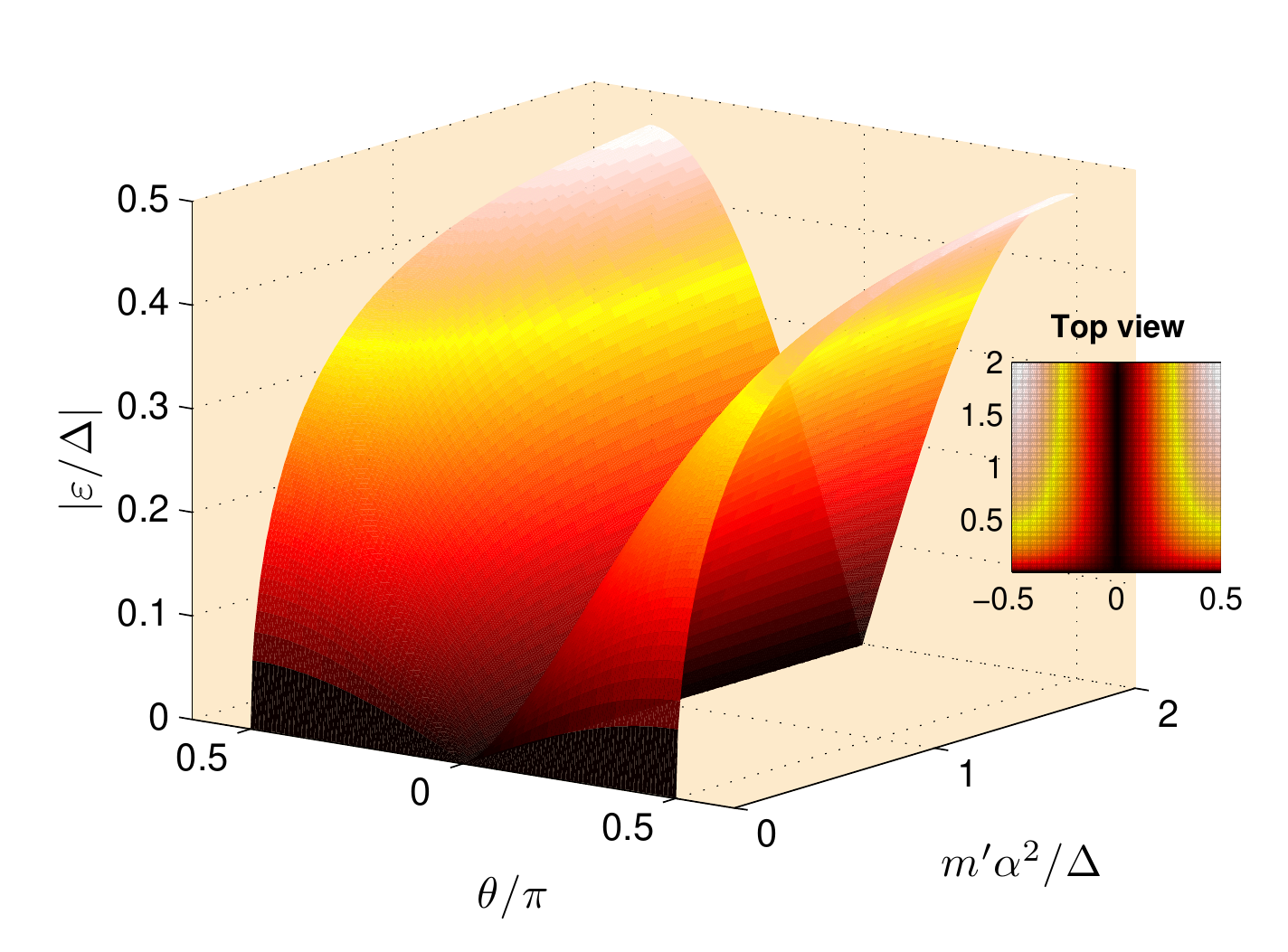}}
\caption{(Color online) Dispersion of the interface-bound state as a function of the angle of incidence $(\theta)$ and the normalized 
spin-orbit coupling strength $(m'\alpha^2/\Delta)$. Here, we have fixed $V_z/\Delta=2$, $\mu/\Delta=3/2$, as should be experimentally 
viable for a proximity-induced gap of size $\Delta=0.5$ meV. The experimental signature of this interface-state would be an enhanced 
subgap DOS, in particular near the Fermi level, compared to the otherwise fully suppressed DOS within the gap in the absence of such 
states.
}
\label{fig:topological} 
\end{figure}

We now analyze the behavior of this interface-state using a realistic set of experimental parameters to identify the relevant energy regime 
where it resides and thus may be probed by \eg STM-measurements. The general requirement for the mapping to the spinless $k_x+\i k_y$-wave 
state is that $V_z$ exceeds $\mu$ in magnitude. In addition, it would be desirable to maximize the Fermi momentum $k_F$ to obtain a large 
normal-state DOS for the benefit of superconducting pairing. Considering Eq. (\ref{eq:wavevector}), it is seen that this can be obtained 
either via a large $V_z$ or large $m'\alpha^2$. The magnitude of $V_z$ will be largely determined by the interface properties (such as 
lattice mismatch) of the ferromagnetic insulator, but values up to a few meV should be within experimental reach \cite{tedrow_prl_86}. 
The spin-orbit coupling strength can to some extent be controlled by a gate voltage, as demonstrated in \eg Ref. \cite{nitta_prl_97}, 
bordering towards 1 K in InGaAs quantum wells. As mentioned previously, the proximity-induced superconducting gap will also be substantially 
reduced compared to its bulk value in the $s$-wave superconductor, and a reasonable estimate would be $\Delta \simeq 0.5$ meV. As a very 
moderate estimate, we then fix $V_z=1$ meV and set $\mu=0.75$ meV; the latter is tunable in a controlled fashion. With these parameters, 
we now plot the interface-state versus the angle of incidence $\theta$ and the normalized spin-orbit coupling strength $m'\alpha^2/\Delta$ 
in Fig. \ref{fig:topological}. As seen, the energy increases with $m'\alpha^2/\Delta$ and eventually saturates around $0.5\Delta$. In this 
plot, we have considered values of $m'\alpha^2/\Delta$ up to 2 in order to demonstrate the evolution of the interface-state in the limit 
of large spin-orbit coupling. Such values may be accessed in a scenario where the proximity-induced gap is very small, \eg $\Delta \leq 0.05$ 
meV (corresponding to a different material choice). For the present choice of parameters, the maximum value of $m'\alpha^2/\Delta$ attainable 
lies around 0.10-0.15. As seen from the plot, the energy of the interface-state is small in this regime, $|\varepsilon/\Delta|\ll1$, and 
reaches zero at normal incidence. This should be readily observable in local DOS measurements at the surface of the quantum well, which 
routinely probe structures with energy-resolution down to $\simeq 200$ $\mu$V \cite{fischer_rmp_07}.

So far, we have established the presence of interface-bound states in semiconducting hybrid structures as shown in Fig. \ref{fig:model} by 
utilizing an exact mapping onto a spinless $k_x+\i k_y$ superconductor model in a realistic parameter regime. However, there are certainly
experimental challenges associated with the proposed structures which we would like to acknowledge here. One point, which in particular 
pertains to the setup in Fig. \ref{fig:model}(a), is related to the Meissner response of the superconductor due to the ferromagnetic 
insulator. This can be avoided by utilizing a ring-like structure (as in Ref. \cite{sau_prl_10}) of the superconducting host material 
which would suppress the orbital effect. In this sense, the structure in Fig. \ref{fig:model}(b) is beneficial since the field here 
resides in the plane of the quantum well, thus strongly suppressing the orbital response. As previously mentioned, another challenge 
is to achieve a sufficiently good interface coupling between the quantum well and the ferromagnetic insulator in order to have an 
appreciable magnitude of the Zeeman-field $V_z$. In this context, we note that EuO has previously been contacted to Al with a 
successfully induced Zeeman-field in Ref. \cite{tedrow_prl_86} as probed by conductance spectroscopy, which demonstrates that such 
a procedure should be feasible. 

In summary, we have investigated an alternative route for experimental observation of Majorana states in semiconducting hybrid structures 
compared to the previously proposed vortex-core states. This route consists of probing interface-bound states via conductance spectroscopy 
or STM-measurements, which we have analytically demonstrated the existence of in this Letter. With a conservative estimate for experimental parameters, we find that these interface-states reside on an energy scale which should be clearly resolvable in such measurements. Whereas 
there are still considerable technological challenges regarding the detection of Majorana fermions in topological insulators, pertaining 
\eg to producing materials of sufficiently high quality, the virtue of the present proposal is that semiconductor technology is very 
well-developed and thus could lead to the experimental observation of Majorana fermions as interface-bound state when utilizing 
present-day methods.

The authors acknowledge support by the Research Council of Norway, Grant No. 167498/V30 (STORFORSK).

\end{document}